# Dispersion-induced generation of higher order transversal modes in singly-resonant optical parametric oscillators

Kai Drühl

Center for Technology Research

Maharishi University of Management

Fairfield IA 52557-1074

e-mail: kdruhl@mum.edu

We study the effects of higher order transversal modes in a model of a singly-resonant OPO, using both numerical solutions and mode expansions including up to two radial modes. The numerical and two-mode solutions predict lower threshold and higher conversion than the single-mode solution at negative dispersion. Relative power in the zero order radial mode ranges from about 88% at positive and small negative dispersion to 48% at larger negative dispersion, with most of the higher mode content in the first mode, and less than 2% in higher modes.

Continuous-wave operation of singly-resonant optical parametric oscillators (SRO) has recently been demonstrated for various non-linear materials, including KTP [1-3] and periodically-poled LiNbO$_3$ (PPLN) [4-6]. Significant pump depletion was achieved, up to 93% in [6]. In these experiments, with zero beam walk-off and confocal parameters of pump and signal comparable to the crystal length, diffractive effects are known to be important [7,8]. Theoretical studies including diffractive effects for cw SROs [7,8] have been published for the case of small pump depletion, and give the dependence of the threshold on the focusing geometry. A treatment of a low-loss SRO for arbitrary pump depletion in the plane-wave approximation was given in [9]. A numerical model including diffraction and pump depletion for a nano-second pulsed SRO with significant birefrigent walk-off was presented in [10]. To the best of our knowledge, no study of diffractive effects for arbitrary pump depletion in a cw SRO without beam walk-off has been published so far. In particular, it is not known how the transverse mode content of the idler beam and the amount of pump depletion depend upon pump power and dispersion. In this letter, we present such results for an SRO with low cavity losses [2]. We find that the idler beam contains higher transverse modes with appreciable amplitudes, and that these amplitudes depend only weakly on the amount of pump depletion, but strongly on the amount of dispersion.

We use the coupled wave equations for signal, idler and pump fields $A_F$ (F=S,I,P) in the form:

$$A_{S,z} = (j/2k_S)(A_{S,xx} + A_{S,yy}) + j\,\kappa_S \exp(j\Delta z) A_P A_I^* \qquad (1.a)$$

$$A_{I,z} = (j/2k_I)(A_{I,xx} + A_{I,yy}) + j\,\kappa_I \exp(j\Delta z) A_P A_S^* , \qquad (1.b)$$

$$A_{P,z} = (j/2k_P)(A_{P,xx} + A_{P,yy}) + j\,\kappa_P \exp(-j\Delta z) A_S A_I . \qquad (1.c)$$

where $\kappa_F = \omega_F\, d/(n_F\, c)$ and d is the effective nonlinear constant. The subscripts x,y and z denote partial derivatives with respect to these spatial coordinates and, for a periodically-poled crystal, $\Delta = k_P - k_S - k_I - 2\pi/l$ is the residual dispersion wave vector ($l$ is the period of the poling). In the limit of small cavity loss $\alpha_S \ll 1$, the signal field has almost constant power. For a Gaussian pump input, the lowest transverse signal mode has largest gain, and therefore reaches threshold first. We found that the threshold for the first higher radial signal mode is about four times higher than for the lowest mode. Our model therefore approximates the signal field by a Gaussian with constant power. In this case, the relationship between pump input power and idler output power is linear, to zeroth order in the cavity loss $\alpha_S$. From photon conservation, the first order increment in signal power is then found to be proportional to the idler output and hence to the pump input. Equating this to the round-trip cavity loss determines the pump input power as a function of signal power [2]. In the limit of small signal power, or small pump depletion, this gives the threshold pump power $P_{thr}$.

In order to understand the influence of dispersion and pump power on the mode content of the idler and pump, it is instructive to consider a mode expansion of the fields $A_F$ into radially symmetric modes $B_{Fn}$ [11,12]. For a zero order signal mode with constant coefficient, the following coupled mode equations result [12]:

$$A_F(x,y,z) = \sum_n a_{Fn}(z) \, B_{Fn}(x,y,z),$$

$$da_{In}/dz = \sum_m c_{nm} \, a_{Pm} \, a_{S0}^*, \qquad (2.a)$$

$$da_{Pm}/dz = \sum_n c_{nm}^* \, a_{In} \, a_{S0}. \qquad (2.b)$$

If pump, signal and idler modes have equal confocal parameter and waist

location, the coupling coefficient $c_{nm}$ takes the form

$$c_{nm} = d_{mn} \exp(j\Phi)/\sqrt{1+(z/z_0)^2}, \tag{3}$$

$$\Phi = z\Delta + (2n+1-2m)\arctan(z/z_0) \cong (\Delta'+1+2n-2m)(z/z_0),$$

$$d_{10} = (1-\lambda_P/\lambda_I) d_{00}, \quad d_{11} = (\lambda_P/\lambda_I) d_{00}, \quad d_{01} = 0.$$

Here $z_0$ is the Raleigh range, $\Delta' = z_0 \Delta$, and the coefficient $d_{00}$ is independent of z. The approximate form of the phase angle $\Phi$ given above is valid for $z/z_0 \ll 1$.

At threshold and with a zero order pump at input (m=0), the effect of the z-dependent phase factor is to favor the n=0 idler mode for $\Delta' = -1$ and the n=1 mode for $\Delta' = -3$. Lowest threshold will occur for $-3 < \Delta' < -1$. The amplitude of the zero idler mode (n=0) will be largest, the amplitude of the first mode (n=1) will be smaller, while the amplitudes of higher modes (n>1) will be suppressed by the larger phase mismatch.

Above threshold, the mode content of the idler will be modified by the higher pump modes (m>0) generated by pump depletion. However, since $d_{0m}=0$, such pump modes are generated only by higher signal modes (n>0). Unless $\Delta'$ assumes large negative values, these have lower amplitude. Furthermore, the coefficients $d_{nm}$, n>0, are small. Thus, we expect the idler mode content to depend only weakly on the pump power.

Our model calculations involve a numerical integration of the coupled wave equations 1.b and 1.c, and of the coupled mode equations 2. The integration of equations 2 uses a second order Runge Kutta algorithm, for the case of zero order modes only (m,n=0; single mode approximation SMA) and for both the zero and first order radial modes (m,n=0,1; two mode approximation TMA).

The integration of 1.b and 1.c uses a split-step algorithm, in which an explicit half-step of parametric conversion is followed by a full step of diffraction and implicit half-step of conversion. The diffraction step uses the Cayley-transform [13] of a discretized version of the radial Laplacian, which conserves photon number exactly. The program was validated by propagating Gaussian modes without conversion, and beams with negligible diffraction and full conversion. Other numerical results were validated by reducing the radial and longitudinal step-sizes by a factor 0.5, giving agreement of better than 0.1% in beam amplitudes.

The models refer to a traveling wave cavity, zero walk-off, equal confocal parameter for pump and signal, crystal length L equal to the confocal parameter $b = 2z_0$ and waist locations at the center of the crystal. Wavelengths are 1.064 µm for the pump and 3.25 µm for the idler. Pump power P and dispersion $\Delta$ are reported as dimensionless parameters P' and $\Delta'=z_0 \Delta$:

$$P = \alpha_S P_0 P',$$

$$P_0 = \varepsilon_0 c\, n_P\, n_S\, n_I\, c^3 / (32 p^2\, \omega_S\, \omega_I\, d^2\, z_0\, k_P).$$

With this unit of power, the threshold parameter $P_{thr}' = P_{thr}/(\alpha_S P_0)$ is related to the threshold function h of [8] by:

$$P_{thr}' = 0.25\, (z_0/L)(1+\lambda_S/\lambda_P)/h\ .$$

For the SMA, the threshold parameter $P_{thr}'$ is given in terms of the Boyd-Kleinman gain reduction factor $h_m$ [7] as:

$$P_{thr}' = 0.5\, (z_0/L)(\lambda_I\, \lambda_S/\lambda_P^2)/h_m\ .$$

Figure 1 shows the threshold parameter $P_{thr}'$ as a function of the dispersion parameter $\Delta'$. The SMA gives minimal threshold at $\Delta'=-0.86$, in good agreement with the discussion above. The TMA and the numerical model give minimal threshold parameter P' for $\Delta'=-1.30$. The TMA agrees very well with the numerical model for the range of $\Delta'$ considered. The SMA agrees with these for $\Delta'>-0.4$, but gives higher threshold parameter for smaller values of $\Delta'$. The minimal threshold parameter is P' = 1.19 for the numerical model and the TMA, and 1.46 for the SMA. The corresponding values for the gain reduction factors h and $h_m$ are h = 0.26 and $h_m$ = 0.778. This agrees with the published values in [8] and [7].

To determine the actual threshold, the wavelength dependence of both the dispersion parameter $\Delta'$ and the resonator loss $\alpha_S$ needs to be considered. The signal wavelength will adjust to a value which gives lowest threshold $P_{thr}$= $\alpha_S P_{thr}' P_0$. If $\alpha_S$ is, within a certain range, independent of wavelength, minimal threshold occurs at a wavelength giving the optimal value of $\Delta'$ = -1.30. If $\alpha_S$ has a minimum at some other wave length, minimal threshold will occur at correspondingly different values of $\Delta'$.

Wavelength acceptance bandwidths for variations in $\Delta'$ have been given in [4] for a PPNL crystal of length L=1cm. For the $d_{33}$ coefficient and a shift in $\Delta'$ of 1.0 above and below the optimal value, the acceptance bandwidth is 7 nm. According to the plane-wave approximation (PWA)[9], such a shift leads to an increase in threshold power by a factor of 1.41. Our numerical model predicts a smaller increase, by a factor of about 1.20.

Figure 2 shows the maximal pump depletion as a function of dispersion. The SMA predicts 100% depletion at $\Delta'=-0.86$ and a pump power of 2.47 times above threshold. This is the same pump power ratio as that predicted by the PWA [9] Both the numerical model and the TMA predict maximal depletion of about 95% at $\Delta'=-1.30$ and a pump power of 2.67 above threshold. This is in good agreement with the experiment reported in [6], which found maximal depletion of about 93% between 2.6 and 3.7 times above threshold. The optimal values of the dispersion parameter $\Delta'$ are the same as those found for minimal threshold. As above, numerical model and TMA agree very well, while the SMA predicts lower conversion and higher pump requirements for $\Delta'<-0.8$. For a shift in $\Delta'$ by 1.0 below and above optimum, our numerical model predicts a reduction of maximal pump depletion to 81% and 75%, while the SMA gives 69% and 68%. The PWA gives 66% in both cases.

These reductions in threshold and increases in pump depletion, as compared to the PWA and SMA, are a consequence of the variable mode structure of the idler beam, as discussed above. As $\Delta'$ is shifted away from the optimal value, any reduction in gain for one of the modes is partially offset by an increase in gain for other modes, leading to a strong dependence of the idler mode content on $\Delta'$.

Figure 3 shows the relative power in the idler modes n=0,1 and n>1 at threshold as functions of $\Delta'$. As predicted, the n=0 mode dominates for small negative and positive values of $\Delta'$, while the n=1 mode becomes increasingly stronger for larger negative values, and dominates for $\Delta'<-2.4$. For the optimal value $\Delta'=-1.3$, the mode n=0 contains 77% of the idler power, the mode n=1 contains 22%, and modes n>1 contain 1%. If the focal point $z=z'z_0$ and the Raleigh range $z_I=z_I'z_0$ of the idler modes are adjusted, to minimize the relative power in modes n>1, such modes contain no more than 2% of the total power. For example, at $\Delta'=-2.3$, for modes with focus at $z'=1.0$ and $z_I'=1.0$, the relative power in modes n=0,1 and n>1 is 51%, 44% and 5%, while for the adjusted modes with focus at $z'=1.04$ and $z_I'=0.74$, the relative power is 65%, 34% and 1%. At $\Delta'=-1.3$, the adjusted modes have focus at $z'=1.04$, Raleigh range $z_I'=0.94$ and contain 80% for n=0, 20% for n=1 and less than 1% for n>1. These values are very close to the values obtained for the original set of modes..

Figure 4 finally shows the idler mode content above threshold as a function of pump power, for adjusted modes, at $\Delta'=-2.3$, -1.3 and -0.3. Fig. 4.a shows the relative power in the mode n=1, fig. 4.b shows the relative phase of modes n=0 and n=1 in degrees, and fig. 4.c shows the location of the mode waist. The relative power in the mode n=1 is seen to be almost independent of pump power, but depends strongly on dispersion. This strong dependence of higher mode amplitudes on the dispersive phase mismatch was also observed in the experimental and numerical studies reported in [10]. The relative phase, which determines the beam radius in the far-field region, vanishes at threshold and increases in absolute value with pump power. This leads to an increase of the far-field beam radius with pump power. The location of the waist shifts towards the entry face of the crystal ($z'=0$), as the pump power increases. The Raleigh range of the adjusted modes is almost independent of pump power, but strongly dependent upon dispersion. It ranges from $z_I'= 0.74$ at threshold to 0.72 at maximal depletion for $\Delta'=-2.3$, from 0.94 to 0.93 for $\Delta'=-1.3$ and from 1.24 to 1.26 at $\Delta'=-0.3$.

In summary, the models studied here demonstrate that the idler output of a cw, low loss SRO with collinear beams contains higher transverse modes with appreciable amplitude, with most of the higher mode content in the first radial mode. The mode content depends strongly on dispersion, but only weakly on pump power. Furthermore, the variable mode content of the idler leads to lower threshold and in higher conversion efficiency for larger negative values of the dispersion parameter $\Delta'$ than those predicted by the single-mode. or plane-wave approximations. The price paid for this is a higher content of radial modes $n>0$ in the idler output.

It is a pleasure to acknowledge Walt Bosenberg at Lightwave Electronics for discussions about the subject of this paper, and for making preprints available.

**Figure Captions**

Figure 1: Threshold pump power as a function of the dispersion parameter $\Delta'=z_0\Delta$, for the numerical, two-mode and single-mode models.

Figure 2: Pump depletion as a function of the dispersion parameter $\Delta'$ at maximal conversion, for the numerical, two-mode and single-mode models.

Figure 3: Relative power in the idler modes $n=0$, $n=1$ and $n>1$ at threshold, as a function of the dispersion parameter $\Delta'$

Figure 4: Mode content of the idler beam as function of normalized pump power $P'=P/P_{thr}$, for $\Delta'=-2.3$, $-1.3$ and $-0.3$. Mode parameters are adjusted for minimal power in modes $n>1$. Fig. 4.a shows the relative power in mode $n=1$, fig. 4.b shows the relative phase of mode $n=1$, and fig. 4.c shows the location $z'=z/z_0$ of the adjusted mode waist.

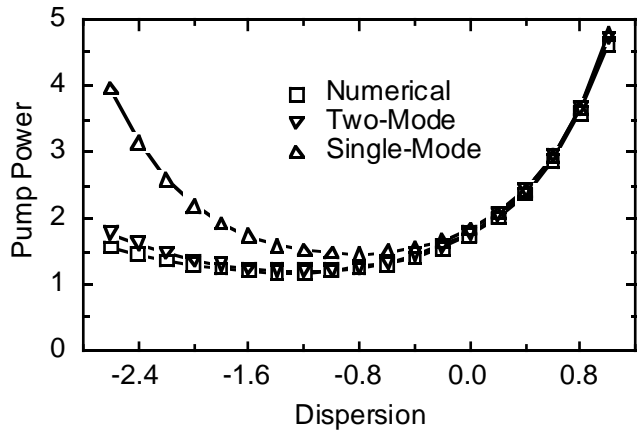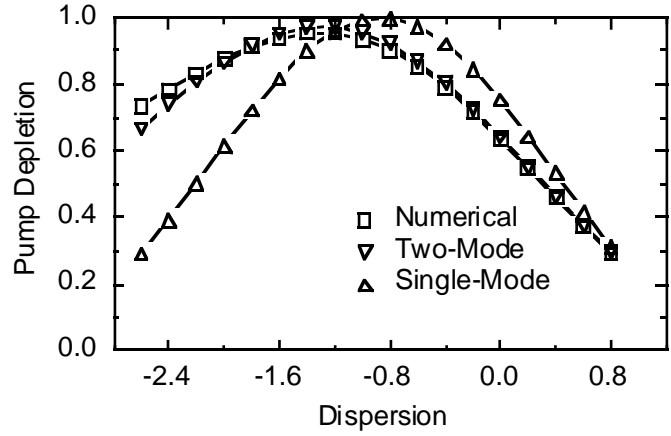

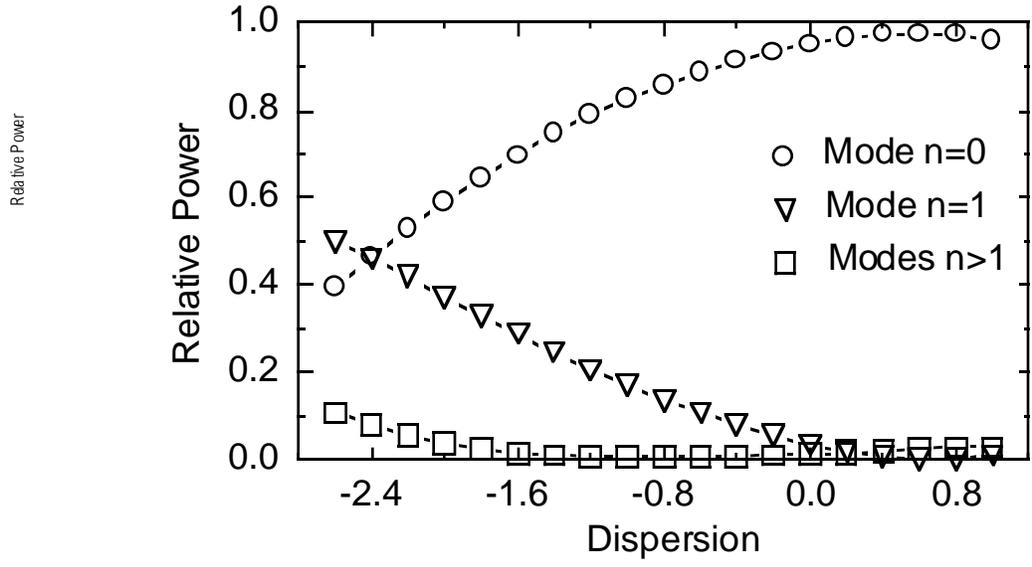

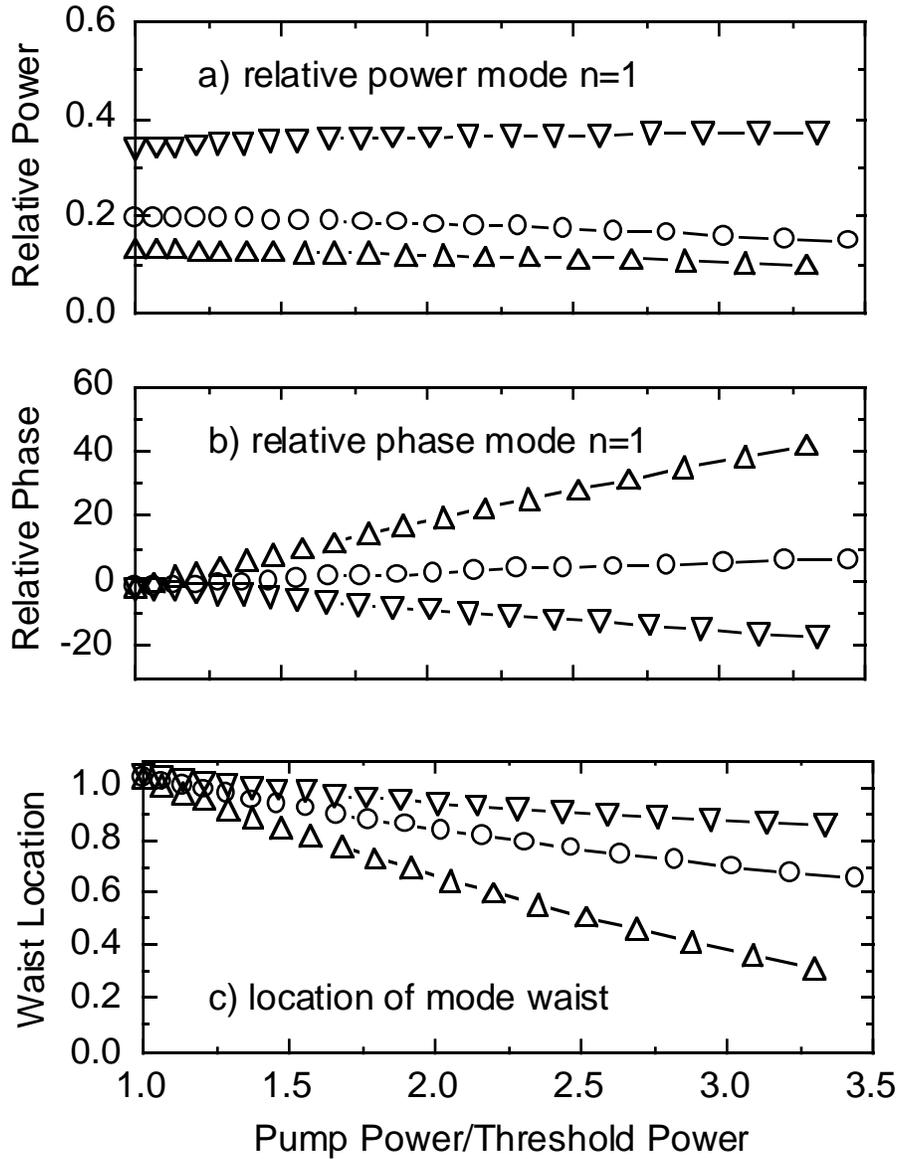